\documentclass[aps,pre,twocolumn,superscriptaddress]{revtex4-1}

\usepackage{epsfig}
\usepackage{graphicx}
\usepackage{dcolumn}
\usepackage{bm}
\usepackage{amsmath,amssymb}
\usepackage{booktabs}
\usepackage{threeparttable}
\usepackage{subfigure,dcolumn}
\usepackage{threeparttable}
\usepackage[normalem]{ulem}

\begin{document}

\title{Machine learning dynamical phase transitions in complex networks}

\author{Qi Ni}
\affiliation{School of Information Science and Technology, East China Normal University, Shanghai, 200241, China}

\author{Ming Tang}\email{tangminghan007@gmail.com}
\affiliation{School of Mathematical Sciences, Shanghai Key Laboratory of PMMP, East China Normal University, Shanghai 200241, China}
\affiliation{Shanghai Key Laboratory of Multidimensional Information Processing, East China Normal University, Shanghai 200241, China}

\author{Ying Liu}
\affiliation{School of Computer Science, Southwest Petroleum University, Chengdu 610500, China}
\affiliation{Big Data Research center, University of Electronic Science and Technology of China, Chengdu 610054, China}

\author{Ying-Cheng Lai}
\affiliation{School of Electrical, Computer and Energy Engineering, Arizona State University, Tempe, AZ 85287, USA}

\date{\today}

\begin{abstract}

Recent years have witnessed a growing interest in using machine learning
to predict and identify critical dynamical phase transitions in physical
systems (e.g., many body quantum systems). The underlying lattice structures
in these applications are generally regular. While machine learning has been
adopted to complex networks, most existing works concern about the structural
properties. To use machine learning to detect phase transitions and accurately
identify the critical transition point associated with dynamical processes on
complex networks thus stands out as an open and significant problem. Here we
develop a framework combining supervised and unsupervised learning,
incorporating proper sampling of training data set. In particular, using
epidemic spreading dynamics on complex networks as a paradigmatic setting,
we start from supervised learning alone and identify situations that degrade
the performance. To overcome the difficulties leads to the idea of exploiting
confusion scheme, effectively a combination of supervised and unsupervised
learning. We demonstrate that the scheme performs well for identifying phase
transitions associated with spreading dynamics on homogeneous networks, but
the performance deteriorates for heterogeneous networks. To strive to meet
this challenge leads to the realization that sampling the training data set
is necessary for heterogeneous networks, and we test two sampling methods:
one based on the hub nodes together with their neighbors and another based on
k-core of the network. The end result is a general machine learning framework
for detecting phase transition and accurately identifying the critical
transition point, which is robust, computationally efficient, and universally
applicable to complex networks of arbitrary size and topology. Extensive
tests using synthetic and empirical networks verify the virtues of the
articulated framework, opening the door to exploiting machine learning for
understanding, detection, prediction, and control of complex dynamical
systems in general.

\end{abstract}

\maketitle

\section{Introduction} \label{sec:intro}

A research frontier across many disciplines of science and engineering is
machine learning~\cite{JM:2015}. In physics, machine learning has attracted
a great deal of attention because of its demonstrated ability to detect,
predict, and uncover various phases of matter in quantum many-body
systems~\cite{Wang:2016,OO:2016,SRN:2017,ZMK:2017,vNLH:2017,CM:2017,CT:2017,
ZK:2017,LQMF:2017,DLD:2017a,DLD:2017b,VKK:2018}. Not only can neural-network
based machine learning generate phases or states of matter that are already
known~\cite{CM:2017,DLD:2017b,ZMK:2017,SRN:2017} or uncover phase
transitions~\cite{Wang:2016,OO:2016,vNLH:2017}, it can also predict
out-of-equilibrium phases of matter that have not been previously
known~\cite{VKK:2018}. In most existing works, the paradigmatic setting
where machine learning, supervised or unsupervised, has been demonstrated to
be effective and powerful is the Ising type of spin dynamics on a lattice.
For example, the principal component analysis (PCA)~\cite{Shlens:2014} for
dimension reduction of data set was exploited to uncover phase transitions
with unsupervised learning, where samples with low and high temperatures
were found to concentrate/distribute in different regions of the learning
space~\cite{Wang:2016}. It was also demonstrated that the threshold or the
critical phase transition point of the Ising model can be predicted through
deep learning~\cite{DL:book}, an important class of machine learning, where
feed-forward neural networks or convolutional neural networks were employed
to extract the necessary structural information~\cite{CM:2017,vNLH:2017}. All
these accomplishments benefited greatly from the regular topology of the
underlying Ising spin lattice.

In the past two decades, researches on complex networks have yielded unprecedented
insights into the working of a large variety of natural, social, and
engineering systems~\cite{Newman:book}. A complex network, by definition,
has a complex topology and, as a natural phenomenon associated with network
dynamics, the emergence of distinct phases and phase transitions are
ubiquitous~\cite{DGM:2008}. The main question to be addressed in this paper
is: \textit{can phase transitions associated with dynamical processes in
complex networks be ``machine-learned''}?

While there were recent efforts in incorporating machine learning into
complex networks~\cite{PAS:2014,GL:2016,KW:2016,WCZ:2016,HYL:2017}, the
studies were limited to learning the \textit{structural information} of
the network. For example, network representation learning in which a complex
network is dimensionally reduced with most of its structure information intact
has found applications in problems such as link prediction~\cite{LK:2007},
clustering~\cite{DHZGS:2001}, and node classification~\cite{BCM:2011}. Based
on random walk and graph search algorithms, the algorithm named ``node2vec''
can embed the network topology into a lower dimensional space by integrating
macroscopic and microscopic structural information about the
network~\cite{GL:2016}. The algorithm ``Deepwalk'' finds a way to unite
skipgram, a natural language processing technique, with random walk sequences
on graphs~\cite{PAS:2014}, providing a concise solution to graph embedding.
Furthermore, it was proved that the performances of deep learning models
such as graph convolutional neural networks~\cite{KW:2016} and structural
deep network embedding~\cite{WCZ:2016} can be equivalent to those of
traditional, random-walk based methods. The key aspect that distinguishes
our present work from the previous works is that we exploit machine learning
to deal with \textit{dynamics} responsible for phase transitions on complex
networks.

To be concrete, we exploit machine learning to predict phase transitions
associated with a fundamental type of dynamics on complex networks:
epidemic spreading~\cite{PSV:2001a,PSV:2001b,Newman:2002,Vespignani:2009,
Vespignani:2012,PSCVV:2015,DGPA:2016,WTSB:2017}. The dynamical process
typically exhibits a second-order phase transition as in the Ising model, and
to accurately identify the threshold or critical point of the phase
transition has been an active research topic. A widely studied method is
the so-called ``degree-based mean field'' approach~\cite{PSV:2001a,DGM:2008},
which gives the theoretical threshold as
$\left\langle k\right\rangle/{\left\langle k^2\right\rangle}$,
where $\left\langle k\right\rangle$ and $\left\langle k^2 \right\rangle$ are
the first and second moments of the degree distribution, respectively.
Strictly, the theoretical prediction is valid only in the limit of infinite
network size, so for any real world networks, there is always a discrepancy
between the predicted and simulated threshold values, where the latter can
be obtained, e.g., by using Monte-Carlo simulations through measures such as
network susceptibility~\cite{FCPS:2012}, variability~\cite{SWTD:2015} or the
average lifetime~\cite{BCPS:2013}. The basic idea and working principle of
our machine learning based approach differ fundamental from those of the
existing methods.

To apply machine learning for predicting the epidemic threshold, a difficulty
must be overcome: a complex network has a hierarchy of structural
irregularities and contains rich features such as hubs, k-cores and
communities, making it challenging for machine learning to grasp the
structural and dynamical information. For example, the PCA method that
is effective for regular lattices usually fails to exhibit any clustering
behavior for complex networks, on which the effectiveness of the learning
algorithm depends. Compounding this difficulty is the complicated interplay
between network structure and dynamics. To meet the challenge, we develop a
systematic learning framework. In particular, we adopt and combine two
different learning methods: supervised and unsupervised learning, to identify
the relation between the configuration data of all nodal states and the
epidemic phase of the system. We demonstrate that, while supervised learning
works well in ideal cases, there is lack of robustness in identifying the
threshold when some labeling information about the training data set is
incorrect or missing. The unsupervised learning method we adopt is
\textit{confusion scheme}~\cite{vNLH:2017}, which can be used to identify
the threshold without requiring any prior knowledge about the labels.
We find that, while this scheme works well for homogeneous networks, it
is largely ineffective for heterogeneous networks. The origin of this
difficulty can be understood by a physical analysis of the relative roles
of the hub nodes and the small-degree nodes in the spreading dynamics.
Aided by this understanding, we articulate two distinct sampling methods
to render the confusion scheme applicable to heterogeneous networks:
\textit{hub-and-neighbors} and \textit{max-k-core} sampling. We show that
incorporating either sampling method can greatly improve the performance of
deep learning in identifying the epidemic threshold for heterogeneous
networks. For example, with sampling the algorithm underlying the confusion
scheme is robust against noise and asymmetry of labeling information.

Overall, our deep learning framework is effective for different types of
network topology, is robust, and is computationally efficient and
consequently applicable to large networks. Our work has thus demonstrated
that machine learning can be powerful for identifying the phase transition
associated with epidemic \textit{dynamics on complex networks of any topology}.
Our framework combining supervised and unsupervised learning as well as
incorporating sampling goes beyond the existing works on learning based
identification of matter phases in regular lattices in physics and those
applicable only to detecting the structural information of complex networks
in computer science. In fact, our deep learning framework provides the base
for potential generalization to broad applications such as predicting the
phases for more diverse types of dynamical processes on complex networks
and identifying the matter phases in complex physical materials.

\section{Epidemic spreading and deep learning framework} \label{sec:definition}

We describe the basics of our machine learning framework for identifying
phase transitions on complex networks, which include the epidemic spreading
model, the structure of the training set, and the neural network model
underlying deep learning.

\begin{figure*}
\centering
\epsfig{file=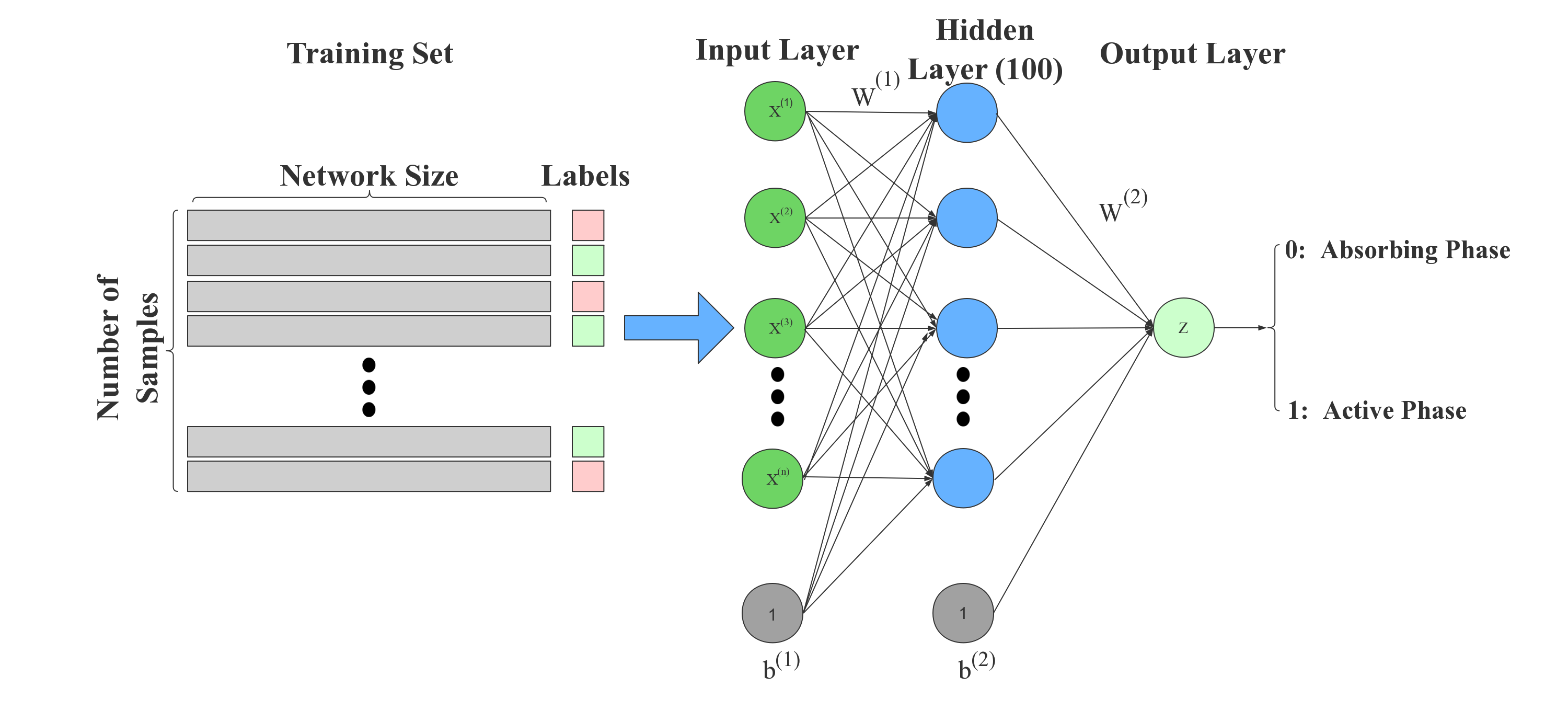,width=1\linewidth}
\caption{ \textit{Illustration of the neural network structure in our deep
learning framework.} Labeled data set is used as input and the learning process
as illustrated is supervised. There is a hidden layer of 100 neurons and a
single-neuron output layer to generate the probability of the identified
phase of the dynamical process on a complex network.}
\label{fig:TF}
\end{figure*}

\subsection{Epidemic spreading and measure of susceptibility}

We consider the classical SIS process on complex networks~\cite{Newman:book}.
During the spreading, each infected node transmits the disease to its
susceptible neighbors at the infection rate $\beta$, and the infected nodes
return to the susceptible state at the recovery rate $\mu$. There are two
distinct phases associated with SIS dynamics: active and absorbing,
where in the former there are both susceptible and infected nodes in the
network but, for the latter, there is no longer any infected node. The
effective infection rate is defined as the ratio of the infection rate to the
recovery rate: $\lambda=\beta/\mu$, and the critical value of the effective
infection rate is denoted as $\lambda_c$. For $\lambda<\lambda_c$, the
system approaches the absorbing state after a long time evolution. In this
case, the system is in the absorbing phase. For $\lambda>\lambda_c$,
asymptotically the system will enter into an endemic state where the density
of the infected nodes reaches a stable value. In this case, the whole
networked dynamical system in the active phase.

We use synchronous updated Monte Carlo method to simulate epidemic
spreading processes in networks. In the absorbing phase,
the nodal states are all identical, rendering them improper for training.
Near the phase transition point, i.e., when the value of $\lambda$ is in the
vicinity of $\lambda_c$ ($\left| \lambda-\lambda_c\right| \agt 0$), there
is a high probability for the system to be trapped in the absorbing state.
To overcome this difficulty, we use the quasi-stationary method~\cite{SCF:2016}
to prevent the system from entering the absorbing state. Especially, whenever
the system tends to the absorbing state, we change its state to that of the
previous time step.

The epidemic threshold of the SIS process can be conveniently characterized
by the measure of susceptibility~\cite{FCPS:2012} defined as
\begin{equation} \label{susceptibility}
\chi=N\frac{\left\langle \rho^2\right\rangle-{\left\langle \rho\right\rangle}^2}{\left\langle \rho\right\rangle},
\end{equation}
where $\rho$ is the density of the infected nodes (i.e., the order parameter),
$N$ is the network size, $\left\langle\rho\right\rangle$ and
$\left\langle\rho\right\rangle^2$ are the first and second moments of $\rho$,
respectively. The order parameter associated with the second-order phase
transition typically exhibits a power-law distribution near the critical
point. As a function of the effective infection rate $\lambda$, the
susceptibility measure reaches its maximum value at $\lambda_c^\chi$, the
threshold of the epidemic process or the phase transition point.

\subsection{Training data set} \label{subsec:training_data}

For the underlying neural network to learn the system dynamics,
proper training data set is needed. To make it more specific,
let us consider that our data set is basically ordered
along a tuning parameter $\lambda$, which is also the control parameter of the
phase transition. The training data with a certain value of $\lambda$ is represented
by $\mathbf{s}(\lambda)$. And it consists of a set of binary vectors,
expressed as $\mathbf{s}^{m\times N}(\lambda)={[a_1,a_2,...,a_N]}^{m\times 1}$,
where $m$ is the number of configurations including
microcosmic dynamical states for all nodes at a time for a given $\lambda$,
and $a_i$ is the binary dynamical state of node $i$ in one configuration.
In particular, $a_i=1$ means that node $i$ is in the infected state
while $a_i=0$ indicates that $i$ is susceptible.
The system phase of every training data of configuration is labeled by function
$l(\lambda)=H(\lambda-\lambda_c)$, where $\lambda_c$ is
the ground truth (or fake truth) that is preset in advance
and $H$ represents the unit step function $H(x)=\frac{d}{dx}max\{x,0\},x\neq 0$.
For any finite size network, we use label $l(\lambda)=0$ to
denote a system configuration being in absorbing state for $\lambda\leq\lambda_c$
and $l(\lambda)=1$ for an active sate with $\lambda>\lambda_c$.

In the training procedure, a large data set of different
$\mathbf{s}(\lambda)$ and corresponding labels $\mathbf{L}(\lambda)$ are fed into the neural
network $\mathcal{F}$. The output of the neural network gradually approaches the preset
label $\mathbf{L}(\lambda)$ through the training process called ``back propagation'' (see appendix
for detailed information). This is to make
$\mathcal{F}(\mathbf{s}(\lambda))\rightarrow \mathbf{L}(\lambda)$ by optimizing the parameter
set of the neural network $\mathbf{P}_\mathcal{F}$ through ``back propagation'' to minimize
a cost function $C(\mathcal{F}(\mathbf{s}(\lambda)),\mathbf{L}(\lambda))$. The cost function is used to
quantify the mismatch between the network's judgment and the real answer. 
It can be minimized by a lot of optimization methods such as
stochastic gradient descent (SGD)~\cite{bottou2012stochastic} and ADAM~\cite{kingma2014adam}.
Actually the neural network is thought to be a high-level mapping function
that takes data $\mathbf{s}(\lambda)$ to infer the probability
distribution $\mathcal{F}(\mathbf{s}(\lambda))=p_0$, where $p_0$ ($1-p_0$)
represents the inferred probability that $\mathbf{s}(\lambda)$ is in the phase of
absorbing (active) state.

\subsection{Neural network architecture}

The gist of our deep learning framework is binary classification in machine
learning, adapted to dynamical processes on complex networks.
The training data set is a set of feature-label pairs
$(\mathbf{s_1},l_1),(\mathbf{s_2},l_2),\dots,(\mathbf{s_m},l_m)$
where $\mathbf{s_i}$ is the feature vector and $l_i$ is the corresponding ground truth
(either $0$ or $1$ to represent two classes) called label.
A classifier $\mathcal{F}$ is supposed to be constructed
by learning the hidden information in the data set,
and is used to classify the new unlabeled data $\mathbf{s_{m+1}}$,
which is called ``test data set''.

The learning process is carried out by a feed forward neural network (FFNN) that
receives a set of labeled training data and passes the data from one layer to
the next - a forward propagation process. The output layer generates results that
can  be compared with the given labels of the training set, triggering a back
propagation process to minimize the cost function which enables the weights
(and biases) of each layer to be updated. More specifically, we construct
our learning model with TensorFlow~\cite{Abadietal:2016}. As illustrated in
Fig.~\ref{fig:TF}, the neural network consists of three dense layers. The
input layer contains $N$ neurons, where $N$ is exactly the size of the complex
network on which the epidemic dynamical process occurs. As the input goes in,
the value in each input neuron is the epidemic state of a node (either 0 or 1).
The hidden layer has 100 neurons, each being fully connected with the input
layer. We impose the ReLU activation function on the hidden layer to achieve a
nonlinear mapping of the input data to the layer~\cite{NH:2010,GBB:2011}. The
general advantage of ReLU is that it can prevent the detrimental phenomenon of
gradient vanishing in learning and expedite the convergent process as compared
with the sigmoid function. We use L2 regularization method to avoid over-fitting.
The output layer has only one neuron whose output is constrained between zero
and one by the sigmoid function, which represents the probability that the
original complex networked system is in a certain phase. Specifically, if the
output is 0 (1), the neural network regards the state of the input data as
belonging to the absorbing (active) phase with probability one. We use Adam
optimizer~\cite{KB:2014} to improve the learning efficiency. Some hyper-
parameters for the neural network are: batch size $N_b=128$, learning rate
$\alpha=0.001$, and regularization parameter $l2=0.01$ (see Appendix for
details).

\section{Identifying threshold through supervised learning and issues}
\label{sec:supervise}

\begin{figure}
\centering
\epsfig{file=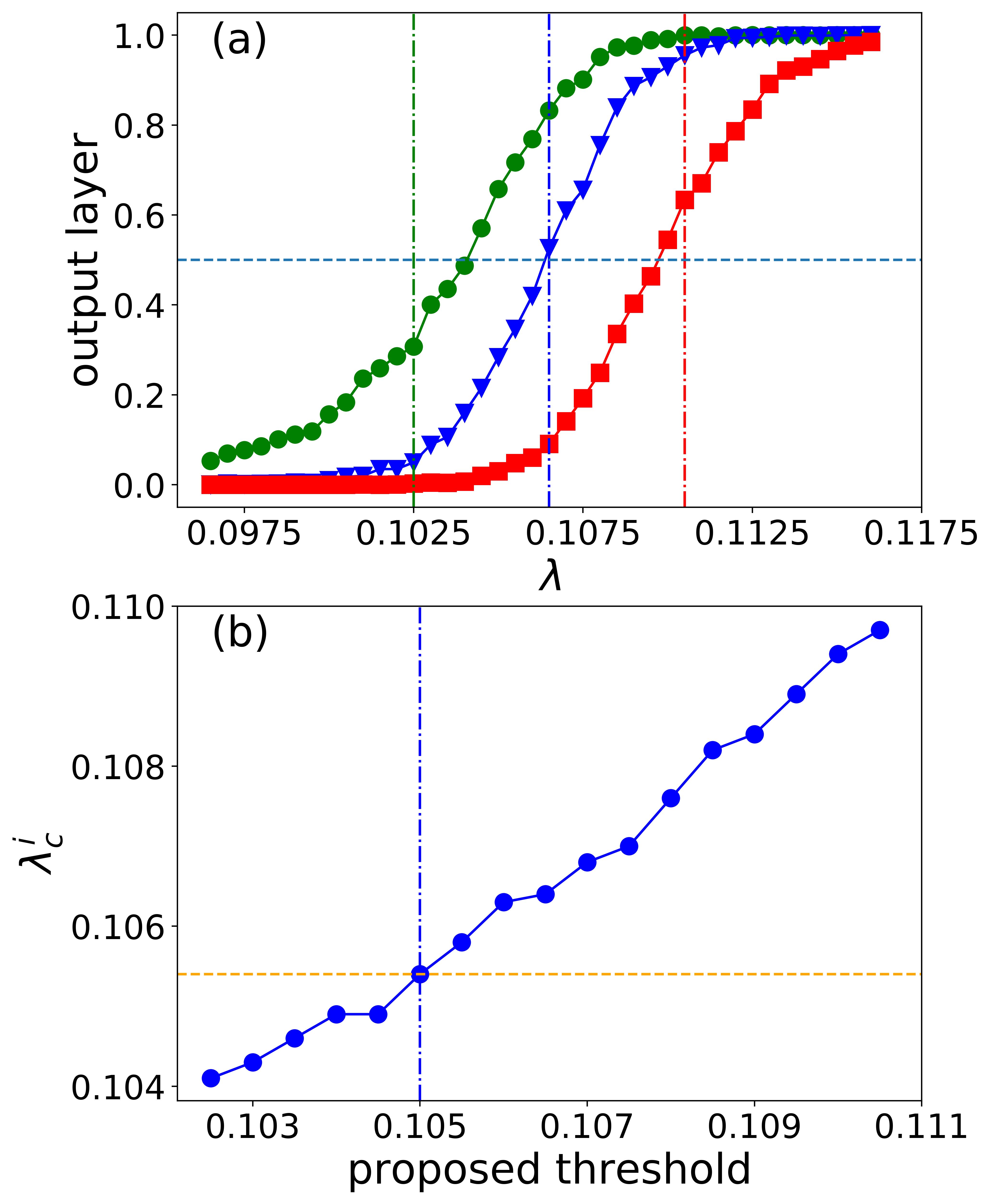,width=1\linewidth}
\caption{ \textit{Identifying phase transition on a random regular network
through supervised learning}. (a) The output of the neural network
averaged over a test set versus the effective infection rate $\lambda$	
for different preset threshold values of the SIS spreading dynamics.
For relatively smaller (larger) value of $\lambda$ than the preset
threshold value as marked by a vertical dashed line, the corresponding
label will be zero (one). (b) The relationship between the identified and
preset threshold values. The neural network's prediction depends on the
preset threshold value. The structural parameters of the random regular
network are $N=1000$ and $\left\langle k\right\rangle=10$.}
\label{fig:RRN}
\end{figure}

The principle to identify the threshold value through supervised learning
can be described, as follows. Suppose a training data set in the matrix
form as described in Sec.~\ref{subsec:training_data} is available, where
each row records the states of all nodes in the network at a given time (i.e.,
a given configuration). Suppose further that the corresponding label vector
characterizing the state of each network configuration is available, as
illustrated in Fig.~\ref{fig:TF}. The aim of supervised learning is to
identify the relation between data and labels. In particular, the training
set is fed into the neural network in Fig.~\ref{fig:TF} to reveal some hidden
patterns in the data. For $\lambda \ll \lambda_c$, after multiple times of
training, the neural network can correctly classify the test set even
without the labels. In this case, there is little confusion for the neural
network to recognize the phase associated with the dynamical process on
the complex network, which can then be correctly recognized. For
$\lambda=\lambda_c$, the average output of the last layer will be about 0.5,
meaning that only half of the data are correctly classified. This corresponds
to the case where the maximum amount of confusion arises, providing a
criterion for the neural network to identify the threshold.

\begin{figure}
\centering
\epsfig{file=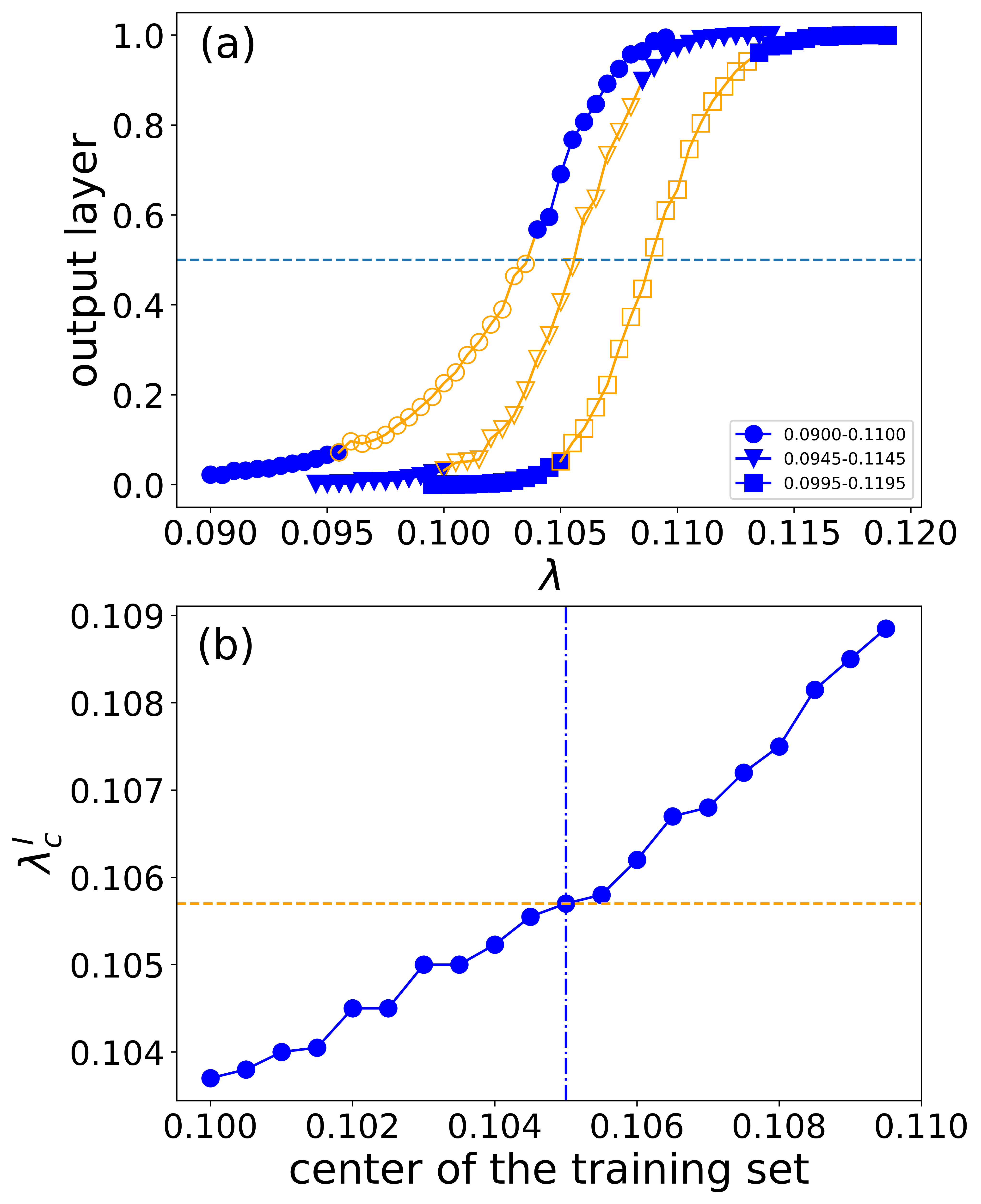,width=1\linewidth}
\caption{ \textit{Identification of phase transition through supervised learning
with truncated training data set on a random regular network}. (a) Threshold
of phase transition identified via supervised learning with truncated data
sets. Data points in the yellow area are artificially removed from the training
set and the neural network makes the judgment only by learning the data in the
blue area. The three curves show that the output shifts when the center
$\lambda$ value of the training set changes from 0.1 to 0.1095. (b) Robustness
against asymmetry of data set. Shown is the relationship between the predicted
threshold value and the center of the training set. When the range of the
training set is shifted while keeping the number of data points in the
training set unchanged, the identified threshold will change.}
\label{fig:TDS}
\end{figure}

To gain insights, we test supervised learning on random regular
networks. Figure~\ref{fig:RRN}(a) shows that, when the output of the
neural network reaches the value of 0.5, the corresponding value of $\lambda$
is quite close to the threshold value $\lambda_c^\chi$. Note that the
identified threshold value depends on the label information in a given
training data set. In an actual situation, we may not know all the label
information of a training set, especially when the state of the underlying
dynamical network is near the threshold. To simulate this situation, we
deliberately change the boundary (i.e., a preset threshold) between the labels
zero and one so as to make incorrect the label information of the training
data between the preset and the true threshold values. For example, there
are wrong labels for the training data between the green and blue dot-dashed
lines in Fig.~\ref{fig:RRN}(a), where the preset threshold is $0.1025$ while
the true threshold is about $0.1050$. As shown in Fig.~\ref{fig:RRN}(b), the
identified threshold value via supervised learning deviates from the true
threshold and increases with the preset threshold value. This means that wrong
labeling information near the threshold can render supervised learning
ineffective at identifying the threshold.

\begin{figure}
\centering
\epsfig{file=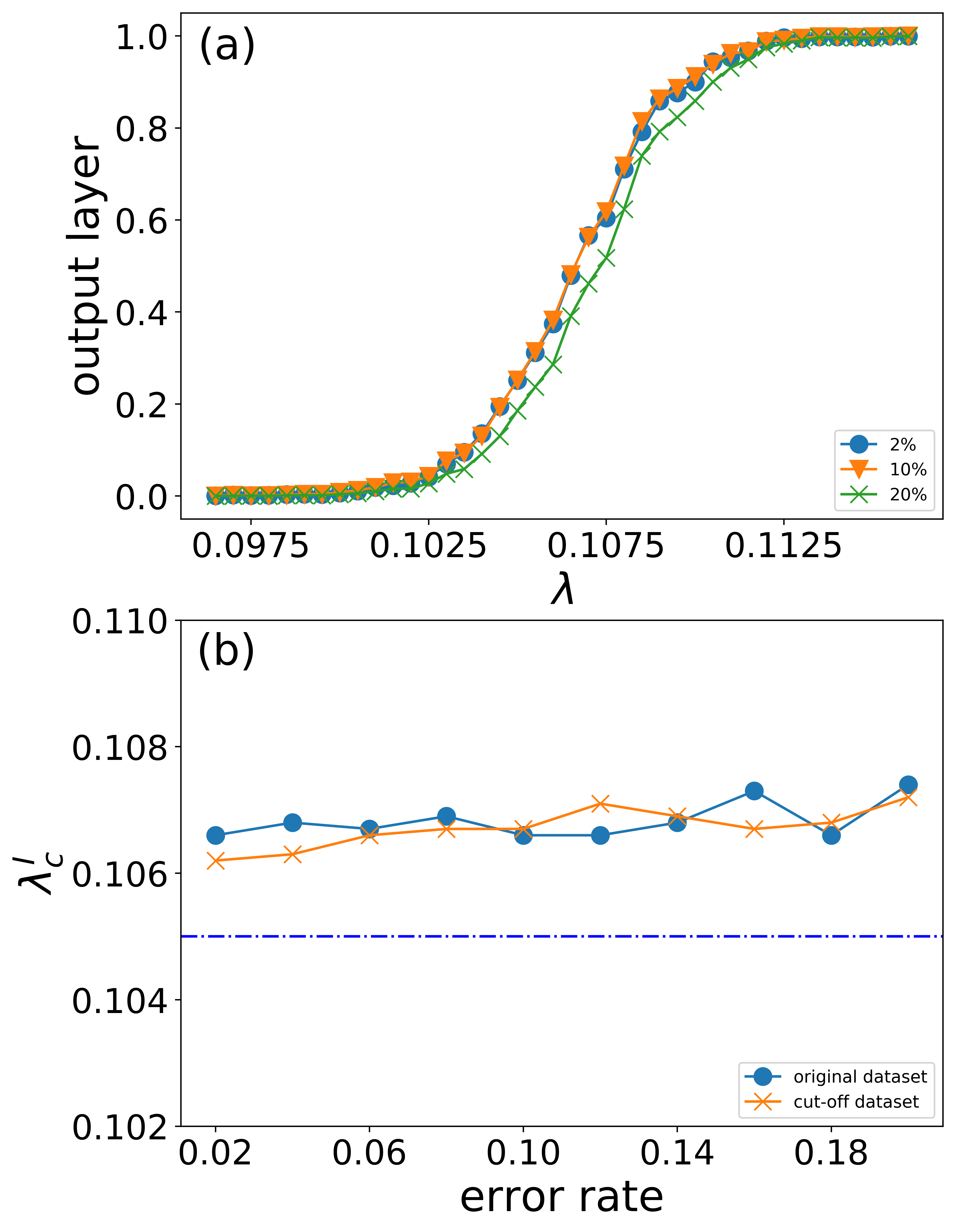,width=1\linewidth}
\caption{ \textit{Robustness of threshold identification via supervised learning
against noise}. For a random regular network, (a) the output curve (output
of the neural network versus the effective infection rate) under different
noisy inputs: noise has little effect on the curve. (b) The relationship
between the identified threshold and the error rate of the training labels.
The predicted value oscillates about the output value with zero noise and
exhibits a slightly upward trend as the labeling error rate is increased.}
\label{fig:noise}
\end{figure}

To overcome the difficulty associated with missing labels and/or the network
spreading dynamics being near the threshold, we truncate the training data
set. In particular, we remove the data points near the critical area and
retain only those far away from the threshold value whose labels are much
less confused. We find that the neural network can still
infer the information pertinent to missing data points by learning the
retained data and identifying the threshold, as shown by the middle curve in
Fig.~\ref{fig:TDS}(a). Another issue that may arise with data set in real
applications is the asymmetry of the truncated data set. To study how the
asymmetry affects the effectiveness of supervised learning, we
shift the data set so that the threshold is no longer the center of the
truncated area. Figure~\ref{fig:TDS}(b) shows that the identified threshold
deviates markedly from the true value, indicating that supervised learning
is sensitive to the asymmetry of the truncated data set.

Is identification of phase transition through supervised learning robust
against noise? To address this issue, we deliberately invert a proportion of
the labels for the training data set and investigate whether the ``artificial''
noise can affect the predicted threshold value. Figures~\ref{fig:noise}(a)
and~\ref{fig:noise}(b) demonstrate that the performance of supervised
learning is robust against noise. Especially, as the labeling error rate
increases, the output of the neural network oscillates but within a
relatively small range.

The results in Figs.~\ref{fig:RRN}-\ref{fig:noise} indicate that supervised
learning for identifying phase transition on random regular networks works
well but only when the data set near the threshold are removed. The
resulting inevitable asymmetry in the truncated data set is detrimental,
but noise has little effect on the performance of supervised learning.

\section{Learning by confusion scheme and the necessity of
incorporating sampling} \label{sec:confusion}

As demonstrated in Sec.~\ref{sec:supervise}, the main deficiency of supervised
learning is that it requires information about the labels to be accurate and
complete and the training data set be symmetric with respect to the true
threshold. To overcome these difficulties, we propose a general framework
combining both supervised and unsupervised learning for accurate and efficient
identification of dynamical phase transition in complex networks. We exploit the
method of confusion scheme that can make precise prediction without any
prior knowledge about the labels~\cite{vNLH:2017}. The confusion scheme is
a supervised learning method utilizing some thoughts of unsupervised learning. The only difference
between confusion scheme and original supervised learning is that the labels
of confusion scheme method are man-made guesses instead of ground truths. As
our data are indexed by parameter $\lambda$, we can avoid using any prior
knowledge by making a ``good guess'' of the true critical point, and the number
of candidates in total is merely $\mathcal{M}+1$ where $\mathcal{M}$ is the number
of different $\lambda$ in our data set. It's affordable to
perform a ``brute force''-like guess because the data set is sorted,
and the result can thus be inferred by the output accuracy of our neural network.

\begin{figure}
\centering
\epsfig{file=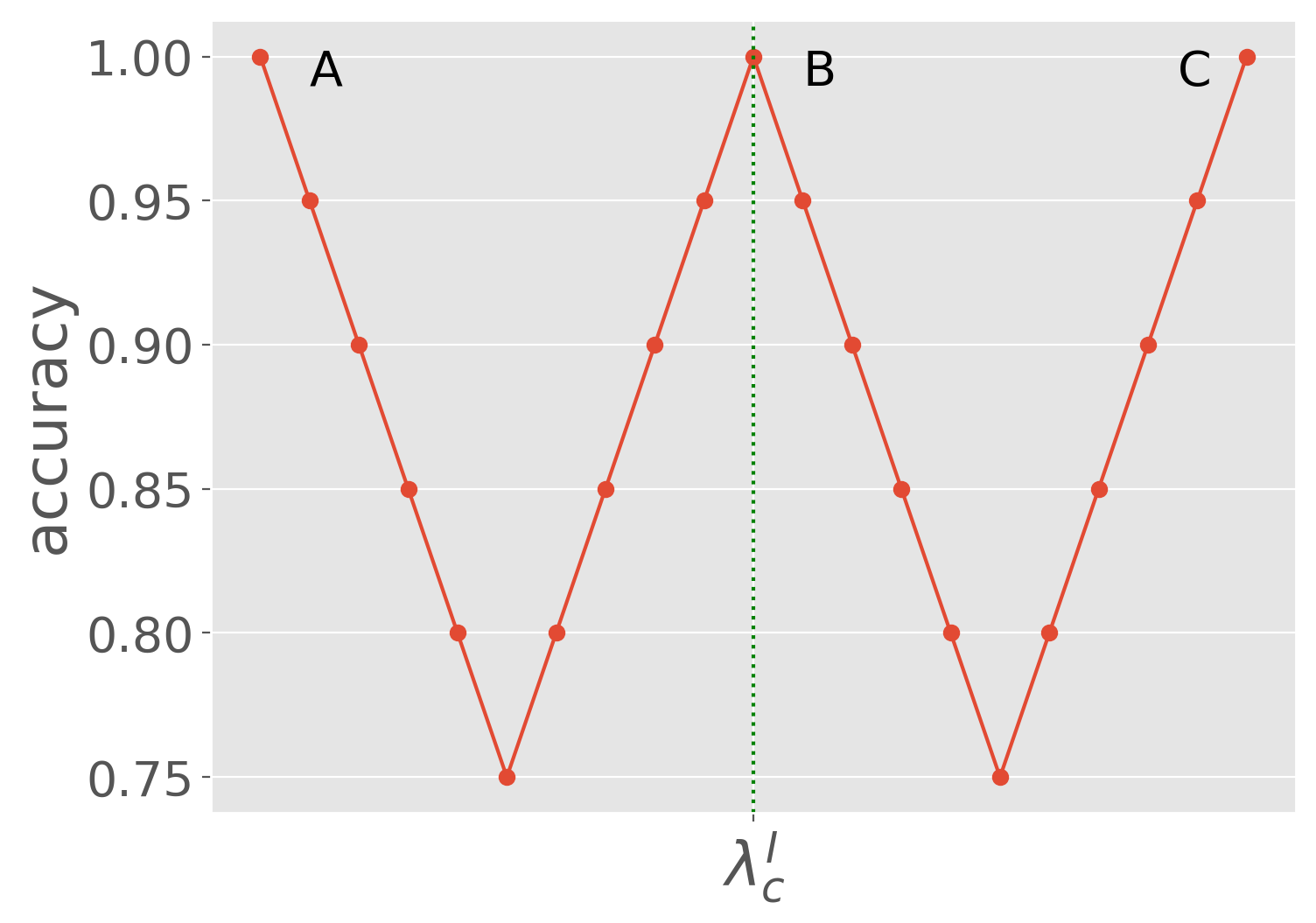,width=1\linewidth}
\caption{ \textit{Schematic illustration of the working of confusion scheme to
detect phase transition and to identify the transition point.} For a range
of tentatively assigned threshold values, the curve of classification
accuracy of supervised learning versus the threshold value will exhibit a
$W$-shape, if there is a phase transition associated with epidemic dynamics
on the network. The location of the middle peak gives an accurate prediction
of the true threshold value. If the network dynamics do not exhibit a phase
transition, the curve would exhibit a $U$ shape (see text for a detailed
reasoning).}
\label{fig:W}
\end{figure}

Suppose we have a set of unlabeled network configuration data ranging from
$\lambda_{min}$ to $\lambda_{max}$. Let $\lambda_c$ be the unknown true
threshold value, where $\lambda_{min}\leq\lambda_c\leq\lambda_{max}$. We
assign tentative labels to the data set by assuming that the threshold
is $\lambda_c^{\prime}$, where $\lambda_{min}\leq\lambda_c^{\prime}\leq\lambda_{max}$.
Especially, we assign label zero to all configurations for
$\lambda\leq\lambda_c^{\prime}$ and label one to those with $\lambda>\lambda_c^{\prime}$.
We then choose a number of closely spaced values of $\lambda_c^{\prime}$. For
each value of $\lambda_c^{\prime}$, we conduct the training and obtain the
classification accuracy. Ideally, we expect the accuracy to exhibit a
$W$-shape kind of behavior versus $\lambda_c^{\prime}$, as schematically
illustrated in Fig.~\ref{fig:W}, which can be argued, as follows. For
${\lambda_c}^{\prime}=\lambda_{min}$, every row of the training set is labeled as
one, so the neural network regards the data of every pattern as in the
activation phase, giving rise to $100\%$ accuracy of prediction. Similar
result is expected for $\lambda_c^{\prime}=\lambda_{max}$. For
${\lambda_c}^{\prime}=\lambda_c$, the method reduces to being supervised learning
because the tentative or ``fake'' labels happen to be correct under the
circumstance. As described in Sec.~\ref{sec:supervise}, the neural network
can yield a high classification accuracy in this case. For
$\lambda_{min}<{\lambda_c}^{\prime}<\lambda_c$ or
$\lambda_{c}<{\lambda_c}^{\prime}<\lambda_{max}$, the neural network will be
``confused'' for some data whose labels are exactly opposite to the true
values, thereby leading to a decrease in the accuracy. Overall, a $W$ shape
of the dependence of the accuracy on the value of $\lambda_c^{\prime}$ arises,
where the location of the peak in the middle corresponds to the identified
threshold. If there is no phase transition in the threshold range
$[\lambda_{min},\lambda_{max}]$, the accuracy versus $\lambda_c^{\prime}$ would
exhibit a universal $U$ shape. The emergence of a $W$-shape curve is thus
unequivocal indication that there is phase transition in the system and the
correct transition point (or threshold) can be identified accordingly without
requiring any prior knowledge about the labels.

\begin{figure*}
\centering
\epsfig{file=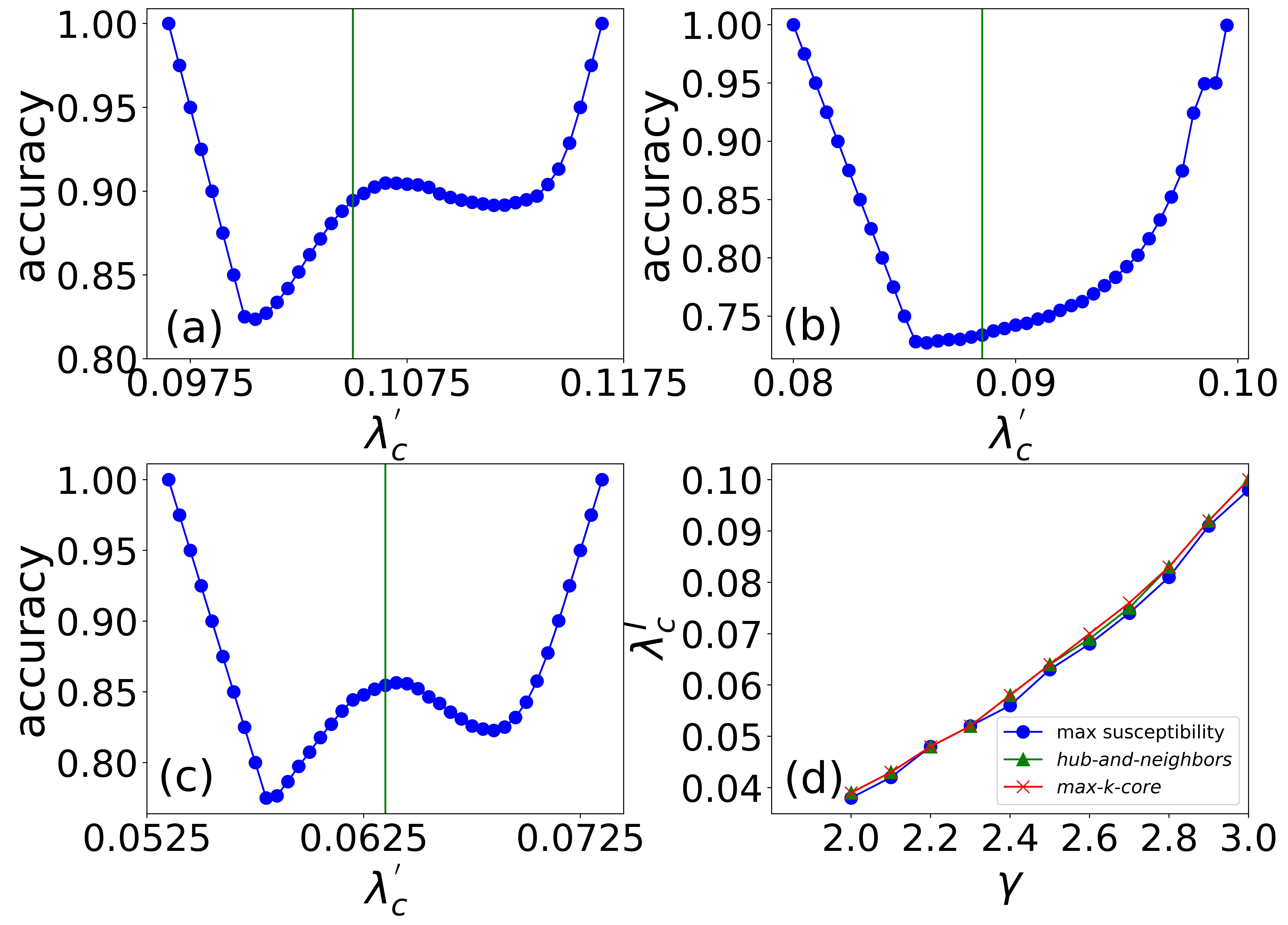,width=1\linewidth}
\caption{ \textit{Learning phase transition by confusion scheme on homogeneous
and heterogeneous networks.} (a) For a random regular network of size $N=1000$
and average degree $\langle k\rangle=10$, the confusion scheme generates an
overall $W$-shape curve of the classification accuracy versus the assumed
threshold value. The peak of susceptibility is at 0.105, as indicated by the
green vertical line (the same legend holds below). (b) For a scale-free
network of size $N=1000$ and structural parameter $m=m_0=3$ generated by the
preferential attachment rule, the neural network fails to yield a $W$-shape
accuracy curve: the accuracy has a $U$-shape. (c) For a scale-free network of
parameters $N=10000$ and $m=m_0=3$, the accuracy curve obtained by
incorporating a hub sampling procedure into the confusion scheme. In this case,
the accuracy curve exhibits a $W$-shape, rendering detectable the phase
transition. The transition point can also be identified accurately. (d) For
scale-free networks generated by the uncorrelated configuration model of size
$N=10000$ with the value of the power law exponent in the range
$\gamma\in[2.0,3.0]$, output of the confusion scheme, where the solid
circles correspond to the maximum susceptibility, the triangles and crosses
indicate the predictions given by neural network incorporating two sampling
methods: $hub$-$and$-$neighbors$ and $max$-$k$-$core$, respectively.}
\label{fig:confusion}
\end{figure*}

We test the confusion scheme on both homogeneous and heterogeneous complex
networks. Figure~\ref{fig:confusion}(a) shows that the scheme can successfully
identify the threshold on random regular networks. For this network topology
and parameters, direct simulation of the SIS dynamics reveals a
second-order phase transition at $\lambda \approx 0.1050$. When the
``fake'' threshold value $\lambda_c^{\prime}$ is varied in a range that contains
the transition point, e.g., $[0.096,0.116]$, the confusion scheme indeed
yields an overall $W$-shape type of behavior and the position of the
middle peak occurs at $\lambda_c^{\prime} \approx 0.1067$, which is
indistinguishable from the actual threshold value. In general, the scheme is
quite effective for detecting phase transition with an accurate identification
of the transition point for homogeneous networks.

For heterogeneous networks such as scale-free networks, a direct application
of the confusion scheme turns out not to be effective. As demonstrated in
Fig.~\ref{fig:confusion}(b), the accuracy curve does not exhibit an apparent
$W$ shape in the chosen threshold range, whereas an actual phase transition
occurs at $\lambda \approx 0.0885$. A possible reason is that, in a typical
heterogeneous network, nodes of relatively large degrees are more prone to
infection. Near the epidemic threshold, there are many nodes of small degrees
which can hardly be infected. These nodes make the data set sparse by assuming
the zero value most of the time and force the neural network to learn with an
unbalanced data set. This will affect the learning process and prevent it
from making the right decision.

\begin{figure*}
\centering
\epsfig{file=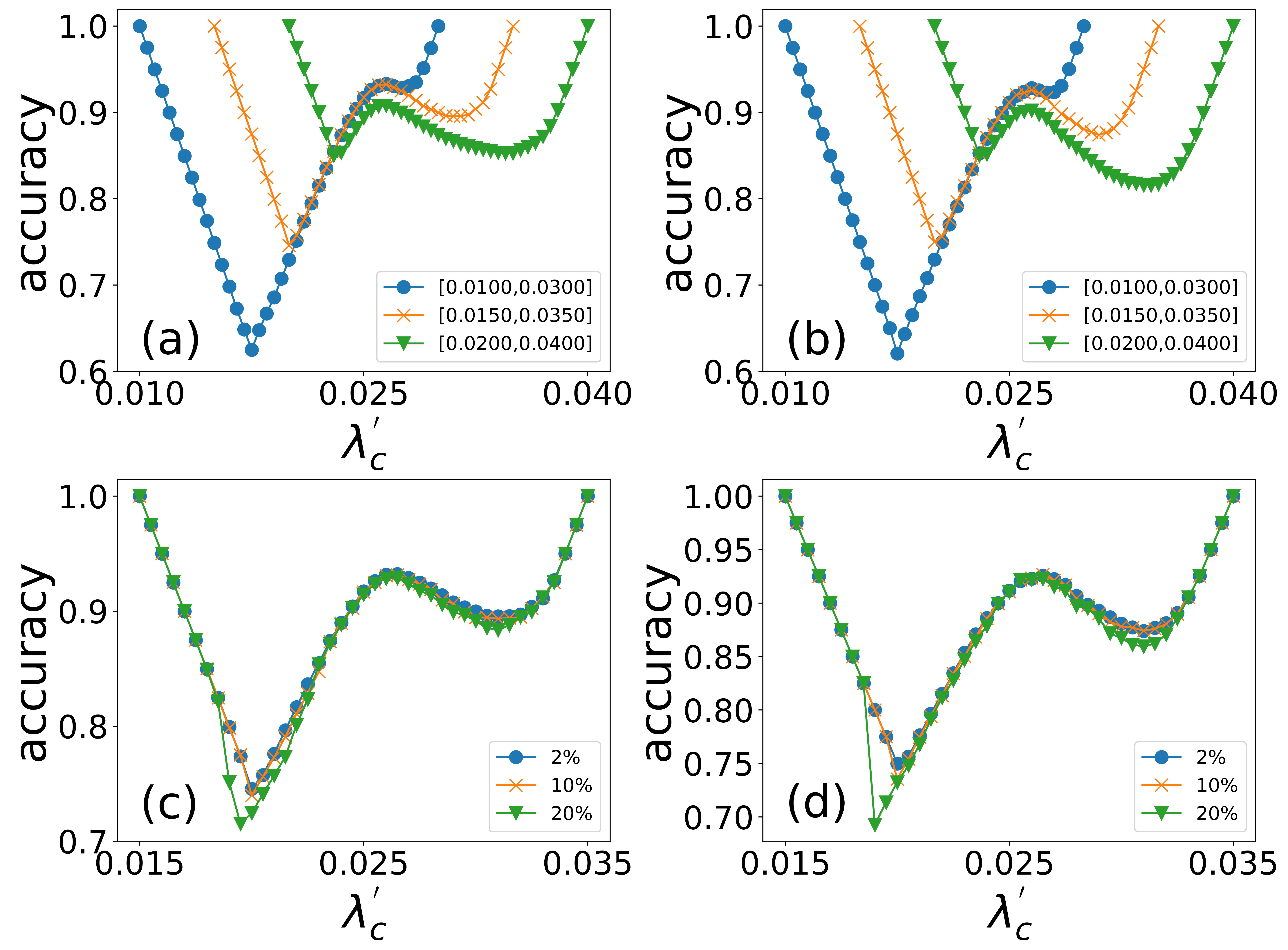,width=1\linewidth}
\caption{ \textit{Performance analysis of the confusion scheme for spreading
dynamics on heterogeneous networks}. The network is scale-free with the
degree exponent $\gamma=2.2$. Symmetry analysis of accuracy incorporating
(a) $hub$-$and$-$neighbors$ and (b) $max$-$k$-$core$ sampling methods.
Different markers indicate different ranges of the training set. (c,d)
Robustness analysis of the two sampling methods. Solid circles, crosses
and triangles represent the results for error rates of $2\%$, $10\%$ and
$20\%$, respectively.}
\label{fig:symmetry}
\end{figure*}

To enhance the applicability of the confusion scheme for
heterogeneous networks, we articulate to incorporate some proper, nodal
importance based sampling procedure into the scheme. Specifically, we aim
to extract the state information of the important nodes and disregard that
from the less important nodes. In the context of epidemic spreading, a
straightforward criterion to determine the nodal importance can be obtained
by addressing the key question: who are the major spreaders? Intuitively,
nodes with large degrees are relatively more important in the spreading
process~\cite{CP:2012}. To provide a physical reasoning, we consider the hub
node with the largest degree in the network. If it is infected, all of its
neighbors are likely to be infected subsequently. However, importance of this
sort will be greatly reduced if the hub node is located at the periphery of
the network, implying that the hub nodes in the central area or core of the
network would have greater importance~\cite{KGHLMSM:2010,LTZD:2015}, where the
max k-core~\cite{BZ:2011} can be used to define the core of the network. We
henceforth propose two sampling methods: (1) to extract the state information
of the hub node with a maximum degree and its neighbors -
$hub$-$and$-$neighbors$ sampling and (2) to extract the information of the
max k-core sub-graph - $max$-$k$-$core$ sampling. As shown in
Fig.~\ref{fig:confusion}(c), when the $hub$-$and$-$neighbors$ sampling
procedure is applied to the same scale-free network in
Fig.~\ref{fig:confusion}(b), the neural network can detect the phase
transition and identify the transition point. For the scale-free network,
there are 314 nodes in the star graph that consists of the hub node and its
neighbors. The peak value of susceptibility is about 0.0635 and the middle
peak of the accuracy curve in Fig.~\ref{fig:confusion}(c) as generated by
the confusion scheme is about 0.0642, which is the predicted transition
point $\lambda_c$. In fact, after incorporating the sampling procedure, the
learning results from the scale-free network are better than those for the
homogeneous network as shown in Fig.~\ref{fig:confusion}(a).

To compare the performances of the two sampling methods, we generate an
ensemble of scale-free networks using the uncorrelated configuration
algorithm~\cite{Newman:book} whose degree exponent ranges from 2.0 to 3.0.
As shown in Fig.~\ref{fig:confusion}(d), both sampling methods perform well,
making the neural network powerful at identifying the epidemic threshold
accurately. As both sampling methods give essentially the same performance,
it suffices to focus on the sampling size (i.e., size of the sub-graph) to
reduce the cost of computation time, which is desired for large networks.

\begin{figure*}
\centering
\epsfig{file=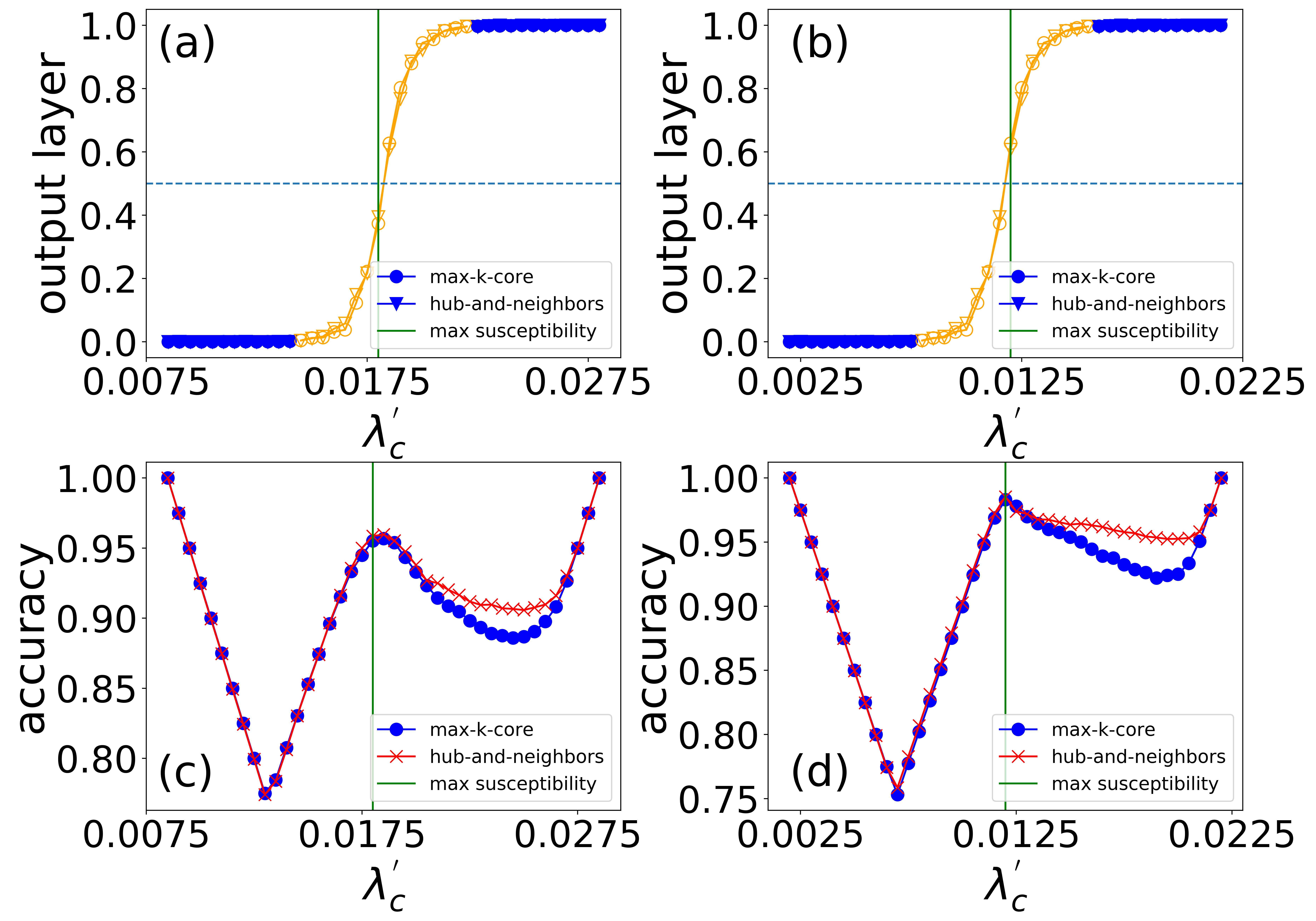,width=1\linewidth}
\caption{ \textit{Test on two representative real-world networks.}
(a,b) Output of the neural network under supervised learning incorporating
$max$-$k$-$core$ sampling for the CAIDA and Brightkite data set, respectively.
Circles and triangles correspond to the results with $max$-$k$-$core$ and
$hub$-$and$-$neighbors$ sampling, respectively. (c,d) Accuracy curves of
executing the confusion scheme for the CAIDA and Brightkite data set,
respectively. For the CAIDA (Brightkite) network,  the size of the sub-graph
sampled is 2628 (1135) with $hub$-$and$-$neighbors$ sampling and 64 (154)
with $max$-$k$-$core$ sampling. Both sampling methods give essentially the
same accuracy result.}
\label{fig:RWN}
\end{figure*}

To address the issue of asymmetric labels associated with supervised
learning, we carry out a symmetry analysis of the confusion scheme. As
shown in Figs.~\ref{fig:symmetry}(a) and \ref{fig:symmetry}(b), the middle
peak of the accuracy curve remains unchanged when the range of the training
data set is shifted, indicating that introducing sampling into the confusion
scheme can overcome the difficulty associated with asymmetry arising when
no sampling is performed. Insofar as there is a phase transition in the
given range, the threshold can be accurately identified.
Figures~\ref{fig:symmetry}(c) and \ref{fig:symmetry}(d) demonstrate the
robustness of the confusion scheme against noise or mislabeling.
As the proportion of the erroneous labels increases from 2$\%$ to 20$\%$,
the accuracy curve exhibits only small fluctuations, and the location of
the middle peak does not change and can still be unequivocally identified.

\section{Applications to real world networks} \label{sec:real-world}

We test the performance of our framework combining supervised learning and
the unsupervised confusion scheme on a number of real world networks.
Figures~\ref{fig:RWN}(a) and \ref{fig:RWN}(b) illustrate the outputs of the
neural network under supervised learning with two real world data set
incorporating the two sampling methods. (From a computational standpoint,
$max$-$k$-$core$ sampling is often preferred because the
$hub$-$and$-$neighbors$ sampling method relies on the hub-and-neighbors star
graphs whose sizes are typically much greater than the max k-core.)
Figures~\ref{fig:RWN}(c) and \ref{fig:RWN}(d) demonstrate the accuracy of
the confusion scheme on the same networks. The accuracies resulting from
the two sampling methods are essentially the same. Performances of our
framework on nine real world networks are summarized in Table~\ref{tab:RW}.
The results with real world networks thus confirm that our deep learning
framework based on combined supervised and unsupervised learning and
incorporating proper sampling is fully capable of ascertaining phase
transition and identifying the transition point associated with spreading
dynamics on complex networks with high accuracy and fidelity.

\begin{table}
\centering
\caption{{\bf Summary of performance results with nine real-world networks}.
The sampling method used is $Max$-$k$-$core$ and the star * indicates
the threshold determined by this method.}
\label{tab:RW}
\begin{tabular*}{8.6cm}{p{1.6cm}<{\centering}p{1.4cm}<{\centering}p{1.8cm}<{\centering}p{1.6cm}<{\centering}p{1.6cm}<{\centering}}
\hline
\hline
Network & Size & Suscept.* & Supervised* & Confusion* \\
\hline
CAIDA & 26475 & 0.0180 & 0.0183 & 0.0185 \\
Brightkite & 58228 & 0.0120 & 0.0118 & 0.0120 \\
Astro-Ph & 18771 & 0.0115 & 0.0117 & 0.0120 \\
PGP & 10680 & 0.0180 & 0.0318 & 0.0315 \\
RV & 6474 & 0.0180 & 0.0310 & 0.0310 \\
Facebook & 4039 & 0.0070 & 0.0070 & 0.0070 \\
Gnutella4 & 10876 & 0.0635 & 0.0643 & 0.0645 \\
Gnutella5 & 8846 & 0.0485 & 0.0497 & 0.0495 \\
Gnutella6 & 8717 & 0.0530 & 0.0534 & 0.0540 \\
\hline
\hline
\end{tabular*}
\end{table}

\section{Discussion} \label{sec:discussion}

To summarize, we have developed a deep learning framework to detect phase
transition and to accurately identify the critical (transition or threshold)
point associated with epidemic spreading dynamics on complex networks. The
main motivations are two-fold. First, in recent years there has been a
great deal of attention in the physics community to exploiting machine
learning for detecting phase transitions in many body quantum systems, but
the lattice structures underlying all existing works in this area are
regular~\cite{Wang:2016,OO:2016,SRN:2017,ZMK:2017,vNLH:2017,CM:2017,CT:2017,
ZK:2017,LQMF:2017,DLD:2017a,DLD:2017b,VKK:2018}, raising the question of
whether deep learning can be effective for identifying phase transitions
in complex networks. Second, while deep learning has been introduced into the
field of complex networks~\cite{PAS:2014,GL:2016,KW:2016,WCZ:2016,HYL:2017},
the existing works dealt exclusively with the structural properties.
Physically, phase transitions in complex networks are often associated with
certain dynamical processes. To apply machine learning to probe dynamical
phase transitions in complex networks in terms of detection, prediction,
and identification was then an unexplored territory, yet the problem is
significant and challenging especially from a physical point of view. Our
work represents an initial effort in addressing this problem.

From the standpoint of methodological development, the innovative aspect of
our framework is a combination of supervised and unsupervised learning,
coupled with sampling methods tailored to complex networks. For a concrete
dynamical process on complex networks, we focus on epidemic spreading.
A straightforward application of supervised learning can be quite successful
in detecting phase transition and predicting the critical threshold,
provided that the labeling information is complete. When there is missing
information about the labels, the performance of supervised learning tends
to deteriorate, often significantly. Truncating the data set to remove
those near the critical point helps to certain extent, but then the inevitable
asymmetry in the data set can make the prediction unstable. To overcome these
difficulties with supervised learning, we exploit the confusion scheme, a
type of unsupervised learning, by which no prior knowledge about the labels
is required. Operated on a systematically chosen set of threshold values,
the confusion scheme constitutes essentially a series of supervised learning
with different assumed knowledge and aims to locate the threshold value that
leads to the maximum amount of ``confusion.'' This combination of supervised
and unsupervised learning performs well and is robust against data asymmetry
and noise, but only for homogeneous networks. For heterogeneous complex
networks, the confusion scheme tends to fail due to the bias in the data
set caused by the intrinsic structure of the network. We find that this
difficulty can be overcome by incorporating proper sampling schemes to
prevent the training data set from being unbalanced. Two sampling methods
have been tested: one based on hub nodes and their neighbors and another
based on k-core - both being quite effective at mitigating the problem of
biased data set for heterogeneous networks. Because the size of the
sampled data set is typically much smaller than the original data size,
sampling has the additional benefit of significantly reducing the computational
load while maintaining the desired accuracy. Through extensive tests
on both synthetic and empirical networks, we conclude that our deep learning
framework is capable of faithfully detecting phase transitions and accurately
pinning down the critical transition point for epidemic spreading dynamics
on complex networks.

Supervised method is a simple, computationally efficient and powerful
solution for identifying critical point. It does need precisely correct
label to support, otherwise its performance might deteriorate.
Nevertheless, it's still useful because supervised method
can classify an one-line input data, i.e., a snapshot
of the nodal dynamical states vector in one time step, into two phases,
which is impossible when using confusion scheme.
To overcome the drawbacks of supervised method,
we can use truncated data set or we can simply let
confusion scheme method generate the good labels first,
and the classification problem would be handled nicely.

\begin{figure}
\centering
\epsfig{file=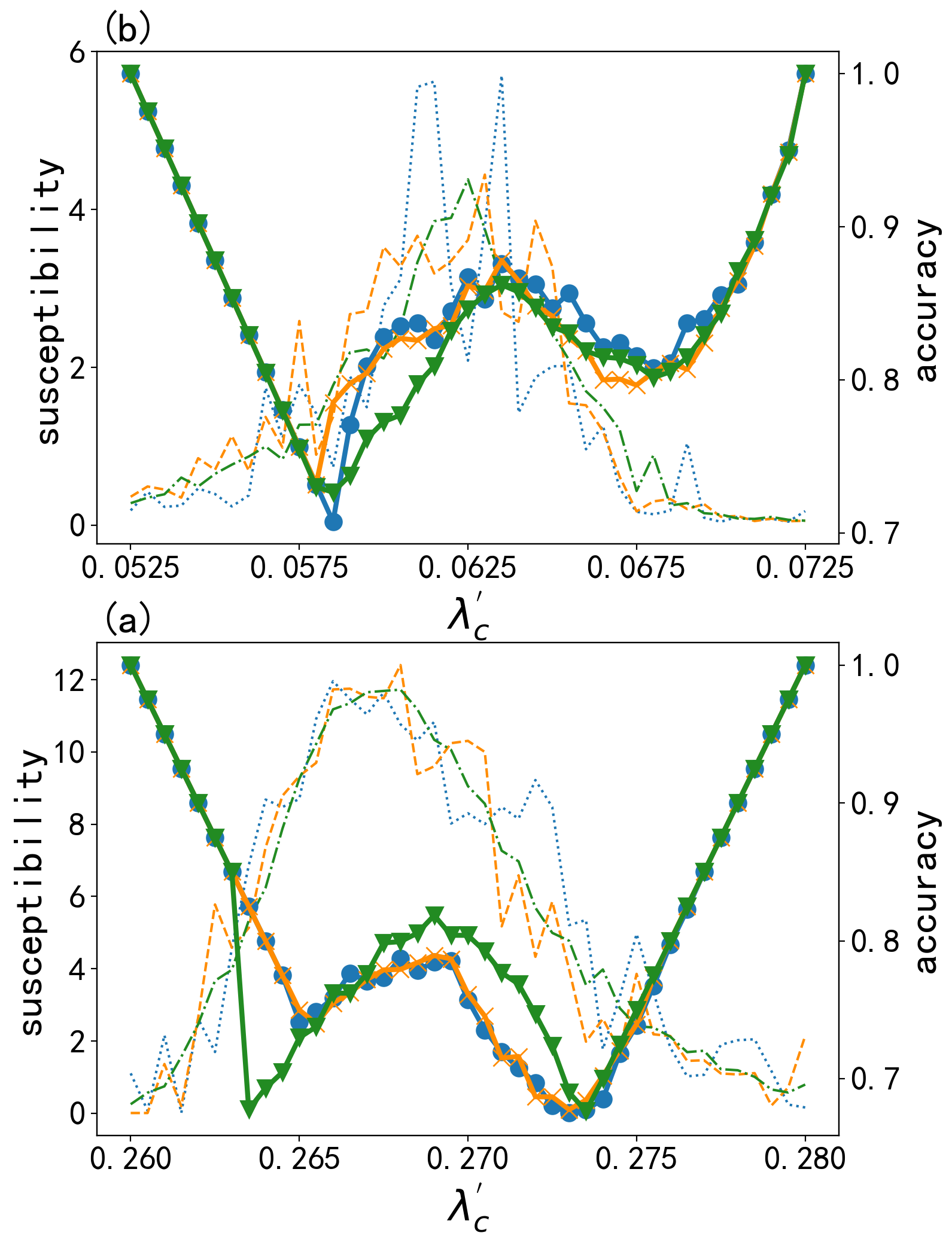,width=1\linewidth}
\caption{\textit{Performance analysis with small data set}. (a) Barabasi-Albert
model with $N=1000$ and $m=m_0=3$.
(b) US power grid network with 4941 nodes.
In each panel, dotted, dashed and dash-dot curves respectively
represent the susceptibility values of data set with $10$, $50$, $100$ realizations for each $\lambda$,
and blue circles, orange crosses and green triangles respectively represent output values of
data set with $10$, $50$, $100$ configurations for each $\lambda$ by using confusion scheme.}
\label{fig:CPR}
\end{figure}

Although the susceptibility measure is simple, precise and
universally applicable in identifying the critical point,
a large number of dynamical configurations are required to
calculate the first and second order moments of $\rho$,
especially near the critical point.
Provided that there are no enough simulations,
the peak of the susceptibility curve will be vague,
and thus the critical point can not be identified accurately.
As shown in Fig.~\ref{fig:CPR}, the susceptibility curve becomes volatile when
the data is not sufficient, while the confusion scheme remains
a relatively stable shape no matter how small the data set is.
Actually, the confusion scheme outperforms the susceptibility measure
almost in every round of comparison due to it's great learning ability.
The confusion scheme requires multiple cycles of supervised learning process, with
every output value in the accuracy curve being independent from one to
another. It is obvious that this multi-round deep learning performs better
than the single-round statistical result, especially when faced with small data set.
When dealing with experiments in real physical system, there are imperfections that
not every labeling information of a physical system is correct or existing,
and thus a deterministic identifying algorithm such as susceptibility may lose its efficacy.
In this case, some machine learning algorithms can automatically adapt to these errors.
Furthermore, our work opens a new direction to use machine learning techniques
in the field of epidemic dynamics in networks.

Comparing with the traditional methods for identifying phase transitions in
complex networks (e.g., numerical approaches based on susceptibility and
other measures), our methods can not only identify the correct transition
point, but classify the given input into different phases as well. Many
open questions remain. For example, can the effect of asymmetrical training
data set on supervised learning be mathematically analyzed? Can certain
optimization methods be developed to increase the accuracy and reduce the
computations in the execution of the confusion scheme? Can our framework
of combining supervised and unsupervised learning be scaled to very large
complex networks? Can a theory be developed to guide the sampling procedure
for heterogeneous networks? Is it possible to develop machine learning
methods to deal with phase transitions in time varying complex networks? etc.
We hope our work will stimulate further efforts in exploiting machine learning
for detecting, decoding, predicting, and even controlling a variety of
dynamical processes on complex networks.

\acknowledgements

This work was supported by the National Natural Science Foundation of China
under Grant Nos.~11575041, 11975099 and 61802321, the Natural Science Foundation
of Shanghai under Grant No.~18ZR1412200, and the Science and Technology Commission
of Shanghai Municipality under Grant No. 18dz2271000. YCL would like to acknowledge
support from the Vannevar Bush Faculty Fellowship program sponsored by the
Basic Research Office of the Assistant Secretary of Defense for Research
and Engineering and funded by the Office of Naval Research through Grant
No.~N00014-16-1-2828.

\section*{Appendix}

Here we list a number of basic notions, concepts, and methods in deep learning
and complex networks, as well as a description of the real world networks
used in our study.

\textit{Feed-forward neural network}, also known as multi-layer
perceptron (MLP), is a basic and powerful deep learning model. The neural
network is structurally dense because each node of the preceding layer is
connected to all nodes in the next layer. The initial layer transmits
information to the next layer through forward propagation, and the process
continues until the output layer. The weights and biases of the whole network
are updated backwards based on descending the cost function along the
gradient to find its global optimum - back propagation. Forward and backward
propagation is executed iteratively until the network finds a global
minimum of the cost function. At this point, the network has successfully
learned the information hidden in the training data set~\cite{RHW:1986,HKP:book}.

What a FFNN does is try to minimize
the cost function which is used to describe the mismatch between the output and
the preset label. Let $w^{l}_{jk}$ be the weight on the edge from the $k$th neuron in
layer $l-1$ to the $j$th neuron in layer $l$ and $b^{l}_{j}$ be the bias on the
$j$th neuron in layer $l$. Thus the input value of the $j$th neuron in layer $l$
is $z^{l}_{j}=\sum_k w^{l}_{jk}a^{l-1}_k+b^{l}_{j}$ and its corresponding output
is $a^{l}_{j}=\sigma(z^{l}_{j})$ where activation function $\sigma$ is non-linear
and differentiable. There's a lot of choices for the cost function such as
quadratic cost $C=\frac{1}{2n}\sum_x(y(x)-a^{L}(x))^{2}$ where $x$, $y$, $a^{L}$
and $L$ represent input data, ground truth labels, output data and maximum number
of layers of the FFNN, respectively. After the optimization target (i.e., cost function) is builded,
we use gradient descend method to minimize it till it reaches the global minimum. Let
$\delta^{l}_{j}$ be the error generated by the $j$th neuron in layer $l$, which
means the deviation between the actual and predicted value. The error in the last
layer can be inferred by the following equations:
\begin{equation} \label{last error}
    \delta^{L}_{j}=\frac{\partial C}{\partial z^{L}_{j}}=\frac{\partial C}{\partial a^{L}_{j}}\cdot \frac{\partial a^{L}_{j}}{\partial z^{L}_{j}}
\end{equation}
or
\begin{equation} \label{last error matrix}
    \delta^{L}=\frac{\partial C}{\partial a^{L}}\odot \frac{\partial a^{L}}{\partial z^{L}}=\nabla_aC\odot\sigma^{\prime}(z^{L}),
\end{equation}
where $\odot$ represents Hadamard product which is an element-wise matrix
product. As the error in the last layer is calculated, errors in other layers can be
obtained recursively as
\begin{equation} \label{other error}
\begin{aligned}
    \delta^{l}_{j}&=\frac{\partial C}{\partial z^{l}_j}=\sum_k\frac{\partial C}{\partial z^{l+1}_{k}}\cdot\frac{\partial z^{l+1}_{k}}{\partial a^{l}_{j}}\cdot\frac{\partial a^{l}_{j}}{\partial z^{l}_{j}}\\
    &=\sum_k\delta^{l+1}_{k}\frac{\partial(w^{l+1}_{kj}a^{l}_{j}+b^{l+1}_{k})}{\partial a^{l}_{j}}\cdot\sigma^{\prime}(z^{l}_{j})\\
    &=\sum_k\delta^{l+1}_{k}\cdot w^{l+1}_{kj}\cdot \sigma^{\prime}(z^{l}_{j})
\end{aligned}
\end{equation}
or
\begin{equation} \label{other error matrix}
    \delta^{l}=((w^{l+1})^{T}\delta^{l+1})\odot\sigma^{\prime}(z^{l}).
\end{equation}
Hence, the gradient of weight $w$ and bias $b$ can be resolved and the gradient
descend method is used to minimize the overall cost function as
\begin{equation} \label{weight gradient}
\begin{aligned}
    \frac{\partial C}{\partial w^{l}_{jk}}&=\frac{\partial C}{\partial z^{l}_{j}}\cdot\frac{\partial z^{l}_{j}}{\partial w^{l}_{jk}}=\delta^{l}_{j}\cdot\frac{\partial(w^{l}_{jk}a^{l-1}{k}+b^{l}_{j})}{\partial w^{l}_{jk}}\\
    &=a^{l-1}_{k}\delta^{l}_{j}
\end{aligned}
\end{equation}
and
\begin{equation} \label{bias gradient}
    \frac{\partial C}{\partial b^{l}_{j}}=\frac{\partial C}{\partial z^{l}_{j}}\cdot\frac{\partial z^{l}_{j}}{\partial b^{l}_{j}}=\delta^{l}_{j}\frac{\partial(w^{l}_{jk}a^{l-1}_{k}+b^{l}_{j})}{\partial b^{l}_{j}}=\delta^{l}_{j}.
\end{equation}

\textit{L2 Regularization}. When training a deep learning model, it is essential
to prevent the model from over-fitting the training set. An over-fitted model
has no practical usage because of lack of generalizability. There are
different ways to deal with the problem such as data augmentation and
regularization. For L2 regularization, a regularization term is added to
the cost function: $C=C_0+ [\lambda/(2n)]\sum_{\omega}\omega^2$,
where $C_0$ is the original cost function,
$[\lambda/(2n)]\sum_{\omega}\omega^2$ is the regularization term, and
$\omega$'s are the weights. We can keep the weights small during the training
process to prevent over-fitting.

\textit{Learning Rate} - an important hyperparameter that controls the speed
at which the weights of the neural network are adjusted based on the loss
gradient. It is a generic parameter in most optimization algorithms such as
SGD, and Adam. The learning rate directly affects how fast the
neural network can converge to a global minimum with the highest possible
accuracy. In general, the greater the learning rate, the faster the neural
network learns. If the learning rate is too small, the network is likely
to fall into some local optimum. However, if the rate is too large and
exceeds some threshold value, the loss of the cost function will oscillate
and stop decreasing.

\textit{Batch size}. If it is not possible to pass data through the neural
network at once, it is necessary to divide the data set into several batches.
Batch size is a hyper-parameter that will affect the training time, which
is essential to finding an optimal value.

\textit{Uncorrelated Configuration Model} (UCM) is a model for generating
complex networks of arbitrary degree distributions~\cite{CBPS:2005}.
In our work, UCM is used to generate random uncorrelated scale-free networks
with a pre-specified degree distribution exponent. The algorithm consists of
two steps. The first step is to assign to each vertex $i$ in a set of $N$
initially disconnected vertices a degree $k_i$, extracted from the probability
distribution $P(k){\sim} k^{-\gamma}$ and subject to the constraints
$m\leq k_i\leq N^{1/2}$ and $\sum_ik_i$ even. The second step is to construct
the network by randomly connecting the vertices with $\sum_ik_i/2$ edges,
respecting the preassigned degrees and avoiding multiple and self-connections.	

\textit{$K$-core decomposition}. For a given complex network, $K$-core is the
sub-graph in which all nodes have at least $K$ neighbors~\cite{Seidman:1983,
KGHLMSM:2010,LTZD:2015}. It characterizes the nodal importance to some
extent. To find the $K$-core of a certain value of $K$, one removes from
the network all nodes of degree less than $K$. Some of the remaining nodes
may have a degree less than $K$ after the removal, in which case one keeps
removing nodes until no node in the core has degree less than $K$. The result,
if exists, is the $K$-core sub-graph.

\textit{Real-world networks}:

\textit{CAIDA}~\cite{LKF:2007}. This is the undirected network of autonomous
systems of the Internet from the CAIDA project, collected in 2007. Nodes
are autonomous systems (AS) and edges represent communication.

\textit{Brightkite}~\cite{CML:2011}. This undirected network contains user-user
friendship relations from Brightkite, a former location-based social network
were users share their locations. A node represents a user and an edge
indicates that a friendship exists between a pair of users.

\textit{Astro$-$Ph}~\cite{LKF:2007}. This is the collaboration network of
authors of scientific papers from the arXiv's Astrophysics (astro-ph)
section. An edge between two authors represents a joint publication.

\textit{PGP (Pretty Good Privacy)}~\cite{BPSDA:2004}. This is the interaction
network of users of the Pretty Good Privacy (PGP) algorithm. The network
has only one giant connected component.

\textit{RV (Route Views)}~\cite{LKF:2007}. This is an undirected network of
the autonomous system of the Internet.

\textit{Facebook}~\cite{LM:2012}. This data set consists of ``circles'' (or
``friend lists'') from Facebook. Facebook data were collected from survey
participants using this Facebook app. The data set includes nodal features
(profiles), circles, and ego networks.

\textit{Gnutella}~\cite{RFI:2002}. This is a sequence of snapshots of the
Gnutella peer-to-peer file sharing network from August 2002. There are
altogether nine snapshots of Gnutella network collected in August 2002.
Nodes represent hosts in the Gnutella network and edges are connections
between the hosts.

\bibliography{DLPTCN}

\end{document}